\documentclass[aps,prb,letterpaper,superscriptaddress,10pt,twocolumn,floatfix]{revtex4-2}

\usepackage[utf8]{inputenc}
\usepackage[T1]{fontenc}
\usepackage{amssymb}
\usepackage{amsmath}
\usepackage{cancel}
\usepackage{graphicx}
\usepackage{dcolumn}
\usepackage{comment}
\graphicspath{{./figs/}}

\begin{document}
\title{Enhanced Emission from Boron-Vacancy Center in Rhombohedral Boron Nitride}

\author{Nasrin Estaji}
\affiliation{Department of Physics, Isfahan University of Technology, Isfahan 84156-83111, Iran}

\author{Ismaeil Abdolhosseini Sarsari}
\affiliation{Department of Physics, Isfahan University of Technology, Isfahan 84156-83111, Iran}

\author{Gerg\H{o} Thiering}
\affiliation{HUN-REN Wigner Research Centre for Physics, P.O. Box 49, H-1525 Budapest, Hungary}

\author{Adam Gali}\email{gali.adam@wigner.hun-ren.hu}
\affiliation{HUN-REN Wigner Research Centre for Physics, P.O. Box 49, H-1525 Budapest, Hungary}
\affiliation{Department of Atomic Physics, Institute of Physics, Budapest University of Technology and Economics, M\H{u}egyetem rakpart 3., H-1111 Budapest, Hungary}
\affiliation{MTA–WFK Lend\"{u}let "Momentum" Semiconductor Nanostructures Research Group, P.O.\ Box 49, H-1525 Budapest, Hungary}

\date{\today}

\begin{abstract}
Boron nitride is a layered crystal whose properties depend on how its atomic sheets are stacked. Its negatively charged boron vacancy is a well-established magnetic defect that can be prepared and read out optically, but in the common hexagonal form it emits very little light, because the symmetry of the surrounding lattice forbids the relevant optical transition. Here we show, using first-principles calculations, that stacking the sheets in the rhombohedral sequence instead removes this restriction and increases the emitted intensity by one to two orders of magnitude, while the magnetic properties remain comparable or improve. We predict that the resulting emission is bright enough for a single defect to be addressed at room temperature, and that a sharp emission line, absent in the hexagonal form, should appear on cooling. Stacking order therefore acts as a design parameter for tailoring the quantum properties of defects embedded in layered materials.
\end{abstract}

\maketitle

\section{Introduction}

Among two-dimensional (2D) semiconductors, boron nitride (BN) stands out for its wide band gap, excellent physicochemical stability, and mechanical robustness. In layered $sp^2$ BN, two stable polytypes are commonly encountered~\cite{olovsson2022rhombohedral, zanfrognini2023distinguishing}: hexagonal boron nitride (hBN) with the $AA^{\prime}$ stacking sequence and rhombohedral boron nitride (rBN) with ABC stacking, which has only recently been demonstrated experimentally~\cite{moret2021rhombohedral, qi2024stacking, iwanski2024revealing}. The stacking sequence strongly influences the optical response and fine structure of BN crystals~\cite{zanfrognini2023distinguishing, iwanski2024revealing}. Light--matter interaction in monolayer BN is intrinsically weak because of the limited interaction length of a single atomic layer, but it strengthens with increasing layer number~\cite{qi2024stacking}. In hBN, inversion symmetry suppresses nonlinear optical interactions between adjacent layers, whereas rBN lacks inversion symmetry and thus naturally supports coherent enhancement of nonlinear optical signals. Accordingly, rBN exhibits distinct advantages, including strong optical nonlinearity, interfacial ferroelectricity, and polarized second-harmonic generation, all arising from its broken centrosymmetry~\cite{wang2024bevel, qi2024stacking}. This symmetry breaking may also enable additional functionalities beyond nonlinear optics.

In the context of quantum technologies, the wide band gaps of 2D BN polytypes make them attractive hosts for color centers. Isolated quantum emitters were initially reported in hBN in the visible~\cite{Tran2016, Shotan2016, Martinez2016} and ultraviolet (UV)~\cite{Bourrellier2016} spectral regions. More recently, it was shown that ABC stacking in rBN can substantially modify the photoluminescence (PL) spectra of UV emitters, and that their spectroscopic fingerprints can be used to identify the rBN host polytype~\cite{iwanski2024revealing}. However, coherent spin control of these UV emitters has not yet been reported, limiting their prospects as qubits for quantum sensing and quantum communication. Here we take a major step toward this goal and show that the lack of inversion symmetry in rBN can be exploited to markedly improve quantum control of embedded, optically addressable defect spins.

Optical initialization and readout of single spins have been reported for several visible quantum emitters in hBN~\cite{Chejanovsky2021, Stern2022room, Guo2023, Patel2024, Stern2024, Gao2025}. However, the microscopic origin of these qubits remains under active investigation. Recent studies suggest that donor--acceptor pairs may underlie some of these spin-active emitters, with random separations between the constituent species~\cite{Auburger2021, Li2025, Robertson2025, Whitefield2026}, which makes deterministic defect engineering challenging. In contrast, the negatively charged boron vacancy (V$_\text{B}^-$) is a well-established optically addressable defect-spin system in hBN~\cite{Abdi2018, gottscholl2020initialization}, and its concentration can be engineered in a controlled manner~\cite{Udvarhelyi2023}. At the same time, coherent control has so far been limited largely to ensembles, because V$_\text{B}^-$ is intrinsically dim and has not been substantially brightened in optical cavities~\cite{Froch2021}. Its extremely low quantum efficiency ($<0.1\%$) originates from the dipole-forbidden nature of the transition between the lowest triplet excited state and the ground state~\cite{reimers2020photoluminescence, ivady2020ab, libbi2022phonon}. Indeed, a recent multireference study of V$_\text{B}^-$ in hBN places the radiative rate of the lowest triplet excited state at only $10.1$~kHz ($\tau_\text{rad}=98.8~\mu$s), whereas the competing spin-selective intersystem crossing (ISC) to the singlet shelving manifold proceeds on a $3.2$~ns timescale~\cite{benedek2026extended}, i.e., the optical cycle is overwhelmingly nonradiative.

In this study, we analyze the photophysics of V$_\text{B}^-$ in 2D BN and show that the lack of inversion symmetry in rBN can enhance its emission intensity relative to hBN. Our \emph{ab initio} results predict a one- to two-orders-of-magnitude increase in the radiative decay rate, with coherent zero-phonon-line emission expected to be observable at cryogenic temperatures in the near-infrared. The simulated room-temperature fluorescence spectrum and the ground-state zero-field-splitting tensor reinforce the tentative assignment of a near-infrared color center to V$_\text{B}^-$ in rBN~\cite{gale2025quantum}. Combined with our previous finding of a long spin--lattice relaxation time $T_1$~\cite{estaji2025spin}, these results establish V$_\text{B}^-$ in rBN as a strong candidate for single-spin quantum sensing. More broadly, our work leverages symmetry engineering to turn a technologically mature defect into a viable qubit platform in rBN that is compatible with cavities, resonators, and nanophotonic components, thereby advancing quantum technologies based on two-dimensional materials.

\section{Results}

\begin{figure*}[!t]
  \includegraphics[width=0.42\textwidth]{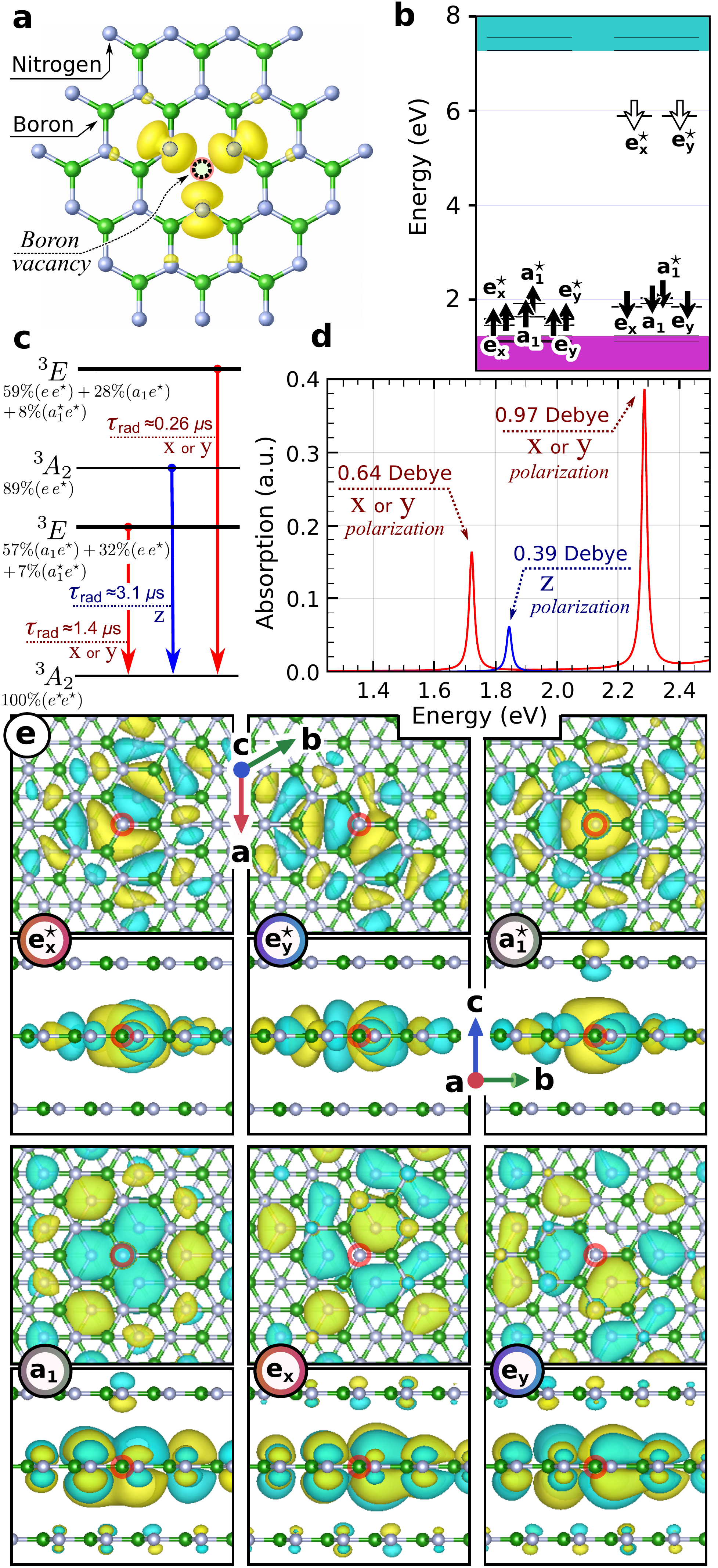}
  \caption{\label{fig:levels-g-e}\textbf{Electronic structure of the negatively charged boron vacancy in rhombohedral boron nitride.}
  \textbf{a} Geometry of the negatively charged boron vacancy with the calculated spin density (yellow isosurface). Nitrogen, boron and the vacant boron site are indicated in the panel.
  \textbf{b} Single-particle Kohn--Sham electronic structure obtained with the Heyd--Scuseria--Ernzerhof (HSE) hybrid functional within density functional theory (DFT). Arrows denote the two spin channels; $a_1$ and $e_{x}$, $e_{y}$ label the fully symmetric and the doubly degenerate defect orbitals, and a star marks the antibonding partners.
  \textbf{c} Many-body electronic structure from many-body perturbation theory within the GW approximation combined with the Bethe--Salpeter equation (GW+BSE). The composition of the exciton wavefunction in terms of single-particle hole--electron pairs is indicated beside the symmetry label of each multiplet excited state. The radiative lifetimes $\tau_\text{rad}$, in microseconds, are derived from the Bethe--Salpeter spectrum.
  \textbf{d} Vertical optical excitation spectrum, with the lowest excitonic peak at 1.72~eV. Red and blue denote photon polarization perpendicular and parallel, respectively, to the threefold rotation axis of the defect. The optical transition dipole moments are given in Debye.
  \textbf{e} Isosurface visualization of the Kohn--Sham orbitals of panel \textbf{b}, in top and side view. The side views show that the mirror plane of the boron nitride sheet is broken, which is what makes the near-infrared transition allowed.}
\end{figure*}

\subsection{Symmetry of the boron vacancy in the two polytypes}

We demonstrate the concept of tailoring the properties of fluorescent defect spins by embedding them in non-centrosymmetric rBN, using V$_\text{B}^-$ as a prototypical example. V$_\text{B}^-$ has a well-defined structure within a single BN honeycomb layer, as shown in Fig.~\ref{fig:levels-g-e}a. In bulk hBN, the AA$^\prime$ stacking enforces $D_{3h}$ symmetry for V$_\text{B}^-$, where the defective layer acts as a mirror plane that effectively plays the role of inversion symmetry~\cite{Abdi2018, ivady2020ab}. Previous theoretical studies~\cite{ivady2020ab, reimers2020photoluminescence} established that the ground state is the $^3A_2^\prime$ triplet, while the lowest triplet excited states have $^3A_1^{\prime\prime}$ and $^3E^{\prime\prime}$ character. Here, the $\Gamma^\prime$ and $\Gamma^{\prime\prime}$ labels denote states that are even and odd with respect to the mirror plane, respectively, in $D_{3h}$ symmetry.

Because electric-dipole transitions between $\Gamma^\prime$ and $\Gamma^{\prime\prime}$ are forbidden in $D_{3h}$ symmetry, the corresponding emission is intrinsically weak and becomes allowed only through vibronic coupling~\cite{libbi2022phonon}. Consistent with coupled phonon--exciton many-body perturbation theory, emission is activated by phonons and coherent zero-phonon-line (ZPL) emission is therefore not expected for V$_\text{B}^-$ in hBN~\cite{libbi2022phonon}. The phonon modes responsible for activating emission involve out-of-plane displacements of atoms surrounding V$_\text{B}^-$, i.e., they break the mirror symmetry~\cite{libbi2022phonon}. This can be rationalized by group theory: out-of-plane vibrations transform as $\Gamma_\mathrm{ph}^{\prime\prime}$ in $D_{3h}$, so the vibronic excited state transforms as $\Gamma_\mathrm{el}^{\prime\prime}\otimes\Gamma_\mathrm{ph}^{\prime\prime}=\Gamma_\mathrm{vibronic}^\prime$, which renders the transition to the $\Gamma^\prime$ ground state allowed (a Herzberg--Teller mechanism).

A higher-lying $^3E^\prime$ triplet excited state resides in the visible spectral range and is dipole allowed. This explains why V$_\text{B}^-$ in hBN can be photoexcited with green light and, following rapid relaxation into the lowest triplet excited states ($^3A_1^{\prime\prime}$ and $^3E^{\prime\prime}$), exhibits weak near-infrared emission via phonon-assisted decay~\cite{gottscholl2020initialization, ivady2020ab, reimers2020photoluminescence}. This analysis suggests that breaking the mirror symmetry should enhance optical emission.

Because the ABC stacking of rBN removes the mirror symmetry (Fig.~\ref{fig:levels-g-e}e), we test this principle for V$_\text{B}^-$ in rBN. The defect symmetry is reduced to $C_{3v}$, such that the prime/double-prime parity labels no longer apply and the associated selection rule is relaxed. We therefore proceed to compute the optical and spin properties of V$_\text{B}^-$ in rBN.

\subsection{Optical properties}

For V$_\text{B}^{-}$ in rBN, we find several occupied defect levels of fully symmetric $a_1$ and doubly degenerate $e$ symmetry. In our calculations with the Heyd--Scuseria--Ernzerhof (HSE) hybrid functional within density functional theory (DFT), the highest occupied $e$ level is filled by two electrons, yielding the $m_S=+1$ component of an $S=1$ state. Because two $a_1$ and two $e$ orbitals appear in the band gap, we label them by their energetic ordering in the spin-majority channel; spin polarization can reorder these levels in the spin-minority channel. All of these orbitals are localized on the three nitrogen dangling bonds adjacent to the vacancy.

Previous work on V$_\text{B}^{-}$ in hBN~\cite{ivady2020ab} showed that several vacancy-localized orbitals with similar energies contribute to strongly correlated many-body excited states. Motivated by this, we investigate the excited-state manifold using many-body perturbation theory within the GW plus Bethe--Salpeter equation (GW+BSE) framework (Methods), which captures correlated excitations as superpositions of single-particle transitions, an essential aspect for V$_\text{B}^{-}$ in two-dimensional BN~\cite{ivady2020ab}. The GW+BSE calculations resolve the low-energy intra-defect excitations and provide the corresponding excitation energies. They also enable analysis of the character of each excited state via its excitonic composition (Fig.~\ref{fig:levels-g-e}c) and yield the optical transition strengths from the ground state.

Throughout these calculations, the atomic geometry is fixed at the DFT-optimized ground-state structure; the resulting spectrum therefore corresponds to vertical excitation energies and neglects ionic relaxation upon photoexcitation (Fig.~\ref{fig:levels-g-e}d). Consequently, the computed spectrum is not intended for direct quantitative comparison with experiment; instead, it provides a reliable map of the optically accessible transition pathways between the ground state and the relevant excited states.

We find that the lowest excited state can be described as a combination of $a_1 \to e^\star$ and $e\to e^\star$ hole--electron pairs which transforms as $^3E$, see Fig.~\ref{fig:levels-g-e}c for visualization. In stark contrast to V$_\text{B}^{-}$ in hBN with $D_{3h}$ symmetry, this lowest excited state is optically allowed from the ground state with perpendicular polarization of photons in the optical transition with $C_{3v}$ symmetry. The excitation wavelength of the lowest excited state falls in the near-infrared (NIR) region. As can be seen in the side views of Fig.~\ref{fig:levels-g-e}e, the planar symmetry is broken, which is responsible for the observable optical transition in the NIR region. We identify another optically allowed state with parallel polarization of photons also in the NIR region, and finally another state in the visible wavelength region (green color) with perpendicular polarization of photons which has the strongest optical transition dipole moment (see Fig.~\ref{fig:levels-g-e}d). The last optical transition also occurs in hBN for V$_\text{B}^{-}$ and can be used to efficiently excite the defect in hBN. Our calculations imply that V$_\text{B}^{-}$ in rBN can be excited in the NIR region too.

Next we focus on the lowest energy triplet excited state which provides luminescence upon appropriate illumination. According to GW+BSE results, the state has $^3E$ symmetry which is subject to Jahn--Teller distortion. We apply the constrained-occupation ($\Delta$SCF) method within HSE DFT to compute the adiabatic potential energy surface of this excited state. To this end, we promoted an electron from the $a_1$ level to the $e^\star$ level in the spin-minority channel. After the self-consistent field procedure in the $C_{3v}$ symmetry, the resulting hole orbital has broken symmetry and shows a mixture of $a_1$ and $e_x$ characters, which implies a multiplet exciton excited state---similar to the GW+BSE result.

We then allowed the ions to relax to a lower symmetry, which indeed occurs due to the $e\otimes E$ Jahn--Teller (JT) effect~\cite{Gali_2016}. Our simulations reveal substantial JT distortion that lowers the total energy by 0.21~eV in the $^3E$ state. The barrier between different JT distorted configurations is 159~meV.

\begin{figure*}[!t]
  \includegraphics[width=0.99\textwidth]{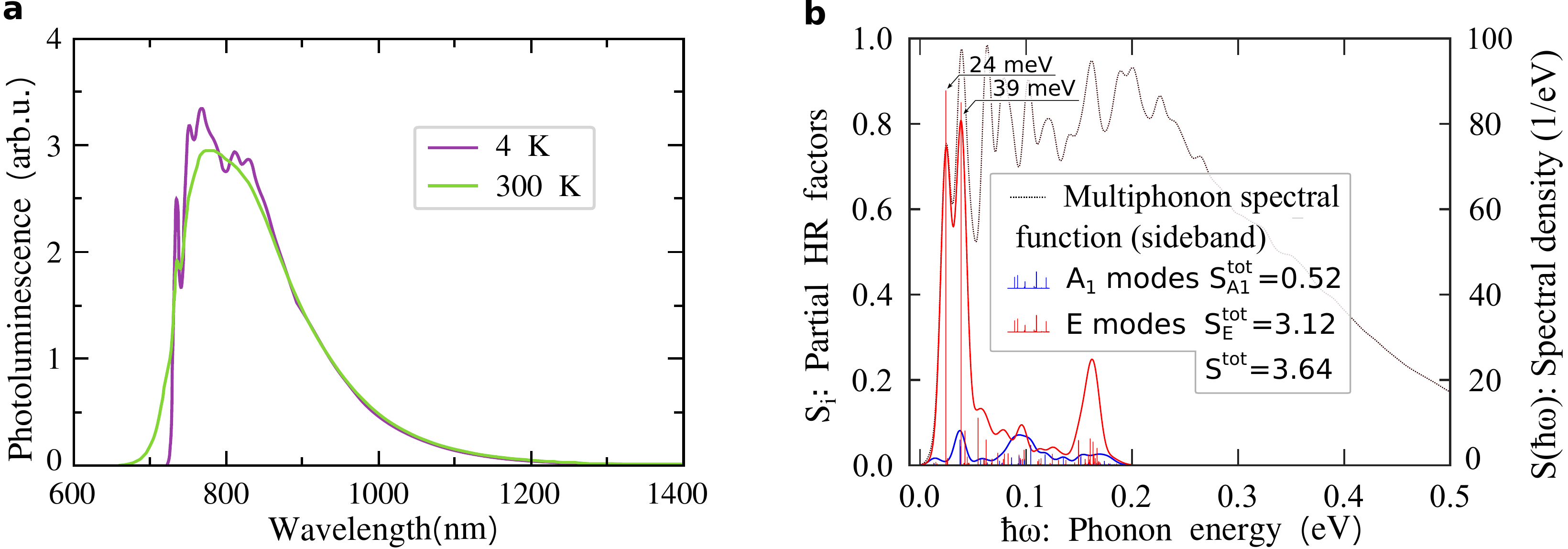}
  \caption{\label{fig:PL}\textbf{Photoluminescence and electron--phonon coupling of the negatively charged boron vacancy in rhombohedral boron nitride.}
  \textbf{a} Photoluminescence (PL) spectrum simulated with the Heyd--Scuseria--Ernzerhof hybrid functional within density functional theory, at 4~K (purple) and 300~K (green). The zero-phonon line (ZPL) and a structured phonon sideband emerge at low temperature. The computed ZPL energy could have an inaccuracy of about 0.1~eV, so the computed spectrum may be shifted when directly compared to experimental data.
  \textbf{b} Electron--phonon coupling underlying the sideband in panel \textbf{a}. Vertical bars are the partial Huang--Rhys (HR) factors $S_i$ of the individual vibrational modes (left axis), resolved into totally symmetric $A_1$ modes (blue) and Jahn--Teller-active $E$ modes (red); the solid curves are the same data broadened by Gaussians for visibility. The HR factors are dimensionless by definition, $S_i=\omega_i q_i^2/2\hbar$, where $\omega_i$ is the frequency and $q_i$ the mass-weighted displacement of mode $i$, hence the left axis carries no unit. The dotted black curve is the multiphonon spectral function $S(\hbar\omega)$ of the sideband (right axis, in eV$^{-1}$), plotted against the phonon energy $\hbar\omega$ (bottom axis, in eV). The two dominant $E$ modes, at 24 and 39~meV, are indicated by arrows. The symmetry-resolved sums given in the legend are $S^{\mathrm{tot}}_{A_1}=\sum_{i\in A_1}S_i=0.52$ and $S^{\mathrm{tot}}_{E}=\sum_{i\in E}S_i=3.12$, giving the total $S^{\mathrm{tot}}=S^{\mathrm{tot}}_{A_1}+S^{\mathrm{tot}}_{E}=3.64$.}
\end{figure*}

The phonon sideband in the luminescence spectrum is determined within the Huang--Rhys (HR) theory~\cite{huang1950theory,Kubo_1955, de_Jong_2015, alkauskas2014first, jin2021photoluminescence, lax1952franck, PhysRevB.88.165202} (see the original theory in Ref.~\citenum{huang1950theory} and our implementation in Ref.~\citenum{thiering2017ab} that is based on Ref.~\citenum{alkauskas2014first}). We calculated the PL spectra of V$_\text{B}^{-}$ in rBN at 4~K and 300~K (Fig.~\ref{fig:PL}a). The defect shows strong room-temperature PL centered at $\lambda\approx734$~nm (1.69~eV). The computed HR factor is $S^{\mathrm{tot}}=3.64$, which results in a relatively broad PL spectrum, but the coherent ZPL peak becomes visible at low temperatures. We note that the sideband originates predominantly from the JT-active modes transforming according to the \(E\) representation of the \(C_{3v}\) group, as indicated by the decomposition of the total HR factor \(S^{\mathrm{tot}}=\sum_i S_i\) into \(E\) (JT-active) and \(A_1\) (totally symmetric) contributions, with \(S^{\mathrm{tot}}_{A_1}=0.52\) and \(S^{\mathrm{tot}}_{E}=3.12\), respectively (see Supplementary Figure~1 in Supplementary Note~1 for details). The resulting Debye--Waller factor, $f_\text{DW}=e^{-S^{\mathrm{tot}}}=2.6\%$, is about six times larger than the $0.4\%$ reported for V$_\text{B}^{-}$ in hBN~\cite{benedek2026extended}, for which the ZPL has never been unambiguously resolved. The computed PL spectrum agrees well with an observed one at room temperature~\cite{gale2025quantum} that was associated with V$_\text{B}^{-}$ in rBN.

The corresponding radiative lifetime ($\tau_{\text{rad}}$) is $1.4~\mu$s, i.e., on the order of a microsecond, estimated as
\begin{equation}
    \Gamma_{\text{rad}}=\frac{1}{\tau_{\text{rad}}}=\frac{n_D E_\text{ZPL}^3 \mu^2}{3 \pi \varepsilon_0 c^3 \hbar^4} \text{,}
\end{equation}
where $\varepsilon_0$ is the vacuum permittivity, $\hbar$ is the reduced Planck constant, $c$ is the speed of light, $n_D=2.1$ is the refractive index of rBN at the ZPL energy $E_\text{ZPL}$, and $\mu=0.64$~Debye is the optical transition dipole moment of the given transition at the GW+BSE level of theory (Fig.~\ref{fig:levels-g-e}d). In hBN, the corresponding emission is symmetry-forbidden at the level of the static $D_{3h}$ structure and is rendered only partially allowed by planar-symmetry-breaking phonons and by the JT distortion, whereas the zero-phonon coherent emission remains symmetry-forbidden~\cite{libbi2022phonon}. The theoretical estimates of the hBN radiative lifetime accordingly span $11$--$99$~$\mu$s, from $\sim11~\mu$s used to model time-resolved optically detected magnetic resonance (ODMR) data~\cite{Mathur_2022}, through $20~\mu$s including Herzberg--Teller intensity borrowing~\cite{reimers2020photoluminescence}, to the multireference value $\tau_\text{rad}=98.8~\mu$s (an upper limit, since electron--phonon coupling shortens it)~\cite{benedek2026extended}. In contrast, the lower $C_{3v}$ symmetry of rBN allows a direct emission with a radiative lifetime that is shorter by one to two orders of magnitude. This is the most important finding of our study, which turns V$_\text{B}^{-}$ addressable at the single defect level in rBN.

\subsection{Nonradiative decay and quantum efficiency}

The brightness of a color center is set by the branching ratio between radiative and nonradiative decay, $\eta=\Gamma_\text{rad}/(\Gamma_\text{rad}+\Gamma_\text{nr})$, so the enhancement discussed above is only meaningful if $\Gamma_\text{nr}$ is not enhanced along with it. Two nonradiative channels compete with photon emission from the lowest triplet excited state: (i) direct multiphonon internal conversion to the $^3A_2$ ground state, and (ii) spin-selective ISC into the singlet shelving manifold. Channel (i) is negligible in both polytypes: bridging the $\approx1.7$--$1.9$~eV electronic gap requires the simultaneous emission of $\gtrsim40$ phonons of the dominant $24$--$39$~meV modes, and the corresponding rate is exponentially suppressed~\cite{benedek2026extended}. Channel (ii) therefore dominates, with $\Gamma_\text{ISC}\approx(3.2~\mathrm{ns})^{-1}$ in hBN~\cite{benedek2026extended}, consistent with the measured excited-state decay rates of $1$--$2$~ns$^{-1}$~\cite{Mathur_2022}.

Crucially, the ISC channel is \emph{not} symmetry-forbidden in $D_{3h}$ hBN---it is already fully allowed and fast---so removing the mirror plane cannot switch it on. Its rate is governed by the spin--orbit matrix elements between the triplet excited state and the singlets, by the singlet--triplet energy gap, and by the vibrational overlap, all of which are properties of the three nitrogen dangling bonds within a single BN sheet. The ABC versus AA$^\prime$ registry perturbs these intra-layer quantities only weakly, as evidenced by the near-identical ground-state zero-field splitting ($D=3.44$~GHz here versus $3.45$~GHz measured in both polytypes~\cite{gottscholl2020initialization, gale2025quantum}), by the nearly unchanged PL band position, and by the fact that room-temperature ODMR---a signal that exists only by virtue of the spin-selective ISC---is readily observed for V$_\text{B}^{-}$ in rBN under essentially the same conditions as in hBN~\cite{gale2025quantum}. Adopting $\Gamma_\text{nr}\simeq\Gamma_\text{ISC}$ unchanged between the two polytypes, we obtain $\eta_\text{hBN}\approx3\times10^{-5}$ and $\eta_\text{rBN}\approx2\times10^{-3}$; that is, the quantum efficiency inherits essentially the full radiative enhancement. We emphasize that this estimate is deliberately conservative: it assumes the least favorable case in which none of the nonradiative rate is reduced in rBN. A quantitative \emph{ab initio} treatment of the rBN ISC rates would require a multireference description of the quasi-degenerate singlet--triplet manifold beyond the scope of the present work~\cite{benedek2026extended}, but any such refinement would have to make the ISC in rBN more than two orders of magnitude \emph{faster} than in hBN to offset the radiative gain, which we consider implausible given the weakness of the interlayer perturbation.

\begin{figure}[!t]
  \includegraphics[width=0.95\columnwidth]{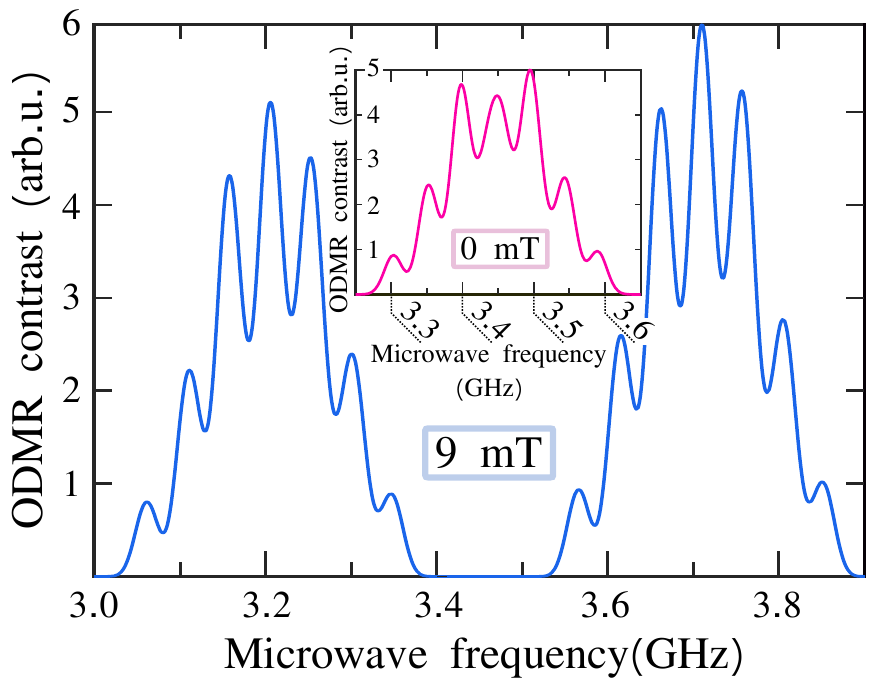}
  \caption{\label{fig:odmr-t1}\textbf{Simulated magnetic resonance spectrum of the negatively charged boron vacancy in rhombohedral boron nitride.}
  Continuous-wave optically detected magnetic resonance (ODMR) spectrum simulated at a magnetic field of 9~mT (blue), computed from the \emph{ab initio} zero-field-splitting and hyperfine tensors reported in this work; the simulation parameters are given in Supplementary Note~4. Inset: the corresponding spectrum at zero magnetic field (magenta), plotted on the same microwave-frequency axis. Both traces show the seven-line hyperfine pattern arising from the three equivalent nearest-neighbor nitrogen-14 nuclei, each of nuclear spin $I=1$, that surround the vacancy. The ODMR contrast is given in arbitrary units.}
\end{figure}

\subsection{Spin properties}

The $^3A_2$ ground state's spin density is localized on the three nitrogen dangling bonds (see Supplementary Figure~2). The respective hyperfine constants are listed in Supplementary Tables~1 and 2. The $C_{3v}$ crystal field splits the $m_S=0$ and $m_S=\pm1$ manifolds of the triplet by dipolar spin--spin interactions. The calculated zero-field splitting (ZFS) parameter is $D=3.44$~GHz, which agrees well with that of a recently observed ODMR center in rBN ($D\approx3.45$~GHz at room temperature) that was associated with V$_\text{B}^{-}$~\cite{gale2025quantum}. We simulated the respective continuous-wave (cw) ODMR spectrum using the calculated hyperfine tensors, which we show at zero magnetic field and at 9~mT in Fig.~\ref{fig:odmr-t1} (see Supplementary Note~4 for details about simulation parameters). Both transitions exhibit resolved hyperfine splitting into seven lines, originating from the nuclear spins ($I=1$) of three equivalent $^{14}$N atoms adjacent to the vacancy in the rBN plane. The nearest-neighbor $^{14}$N nuclei yield a splitting of $A_{zz}\approx48.3$~MHz in the ground state. The hyperfine broadening of the simulated cw-ODMR spectrum and the broadening in the observed cw-ODMR spectrum agree remarkably well, which reaffirms the identification of the ODMR center in rBN.

We note that the reduced symmetry may open additional relaxation channels when compared to those for V$_\text{B}^{-}$ in hBN. The calculated spin--lattice relaxation time for V$_\text{B}^{-}$ in rBN is $T_1\approx25~\mu$s at room temperature, which is analyzed in detail in our previous study~\cite{estaji2025spin}. This indicates that V$_\text{B}^{-}$ in rBN can be applied as a room-temperature qubit.

We further note that by solving the $e\otimes E$ JT Hamiltonian~\cite{thiering2017ab} for the $^3E$ excited state with the use of our developed code~\cite{Toth2025}, we obtained a Ham reduction factor of 0.006, implying strong quenching of spin--orbit coupling and off-diagonal ZFS tensor components~\cite{Thiering2025} due to electron--phonon interactions. As a consequence, the usual $D=3/2\,D_{zz}$ component yields the energy splitting between $m_S=0$ and $m_S=\pm1$ levels in the $^3E$ excited state. On the other hand, the computation of the ZFS tensor is not straightforward with DFT due to the multi-determinant nature of the excited state (see Supplementary Note~3 for details). Nevertheless, we roughly estimate the $D$ constant of the lowest energy $^3E$ state to be approximately $D_{\mathrm{theory}}^{\mathrm{ex.}}\approx 1.5~\mathrm{GHz}$. We note that this \emph{ab initio} value is relatively close to the experimentally observed 2.1~GHz associated with the ZFS of the excited state of V$_\text{B}^-$ in hBN~\cite{Mathur_2022, Baber_2021, Yu_2022, PhysRevLett.128.216402, Gao_2022}. Given the similarity of the ground-state ZFS of V$_\text{B}^-$ in hBN and rBN, our results suggest that the excited-state ZFS of V$_\text{B}^-$ in rBN should also be around 2.1~GHz.

\section{Discussion}

In this work, we have theoretically demonstrated that the rhombohedral polytype of boron nitride provides an environment in which the optical and spin properties of the negatively charged boron-vacancy center are markedly enhanced. By combining density functional theory and many-body perturbation theory, we revealed that symmetry reduction in rBN enables optically allowed transitions with radiative lifetimes on the order of microseconds and photoluminescence that makes the coherent emission visible at low temperatures. The predicted zero-field splitting and hyperfine structure are in excellent agreement with experimental observations, supporting the reliability of our approach.

Our results establish rBN as a promising host material for solid-state spin defects, overcoming the low brightness that has limited the usability of V$_\text{B}^{-}$ centers in hBN. This finding introduces a materials design pathway where layer stacking is not merely a structural detail but a powerful degree of freedom to engineer the quantum properties of embedded defects. Such control opens opportunities for realizing room-temperature single-spin quantum sensors, scalable photonic devices, and hybrid van der Waals quantum technologies.

This enhancement should be kept in perspective, and it has not yet been tested experimentally: the only existing measurement on V$_\text{B}^{-}$ in rBN was not designed as a photometric comparison and reports no defect-density-normalized count rates~\cite{gale2025quantum}. A microsecond radiative lifetime is still three orders of magnitude longer than the nanosecond lifetimes of the bright visible and UV single-photon emitters of BN~\cite{Tran2016, Bourrellier2016}, so V$_\text{B}^{-}$ in rBN remains a comparatively dim emitter. What changes is that it crosses the threshold of single-defect detectability. With the optical cycle limited by the metastable singlet ($\tau_\text{cycle}\approx0.04$--$0.34~\mu$s~\cite{benedek2026extended}) and $\eta_\text{rBN}\approx2\times10^{-3}$, the saturated photon emission rate is $\sim10^4$~s$^{-1}$; assuming a few per cent collection efficiency for a high-numerical-aperture objective, $\sim50\%$ optical throughput and $\sim60\%$ detector quantum efficiency in the near infrared, this corresponds to $\sim10^2$--$10^3$ counts~s$^{-1}$, which is sufficient for photon-correlation measurements on a single center. The same estimate for hBN gives only $\sim1$--$10$ counts~s$^{-1}$, at or below typical background, consistent with the fact that single V$_\text{B}^{-}$ centers have never been isolated optically in that polytype~\cite{gale2025quantum}.

The second consequence is spectroscopic. Because the ZPL is no longer symmetry-forbidden and the Debye--Waller factor rises from $\approx0.4\%$ to $\approx2.6\%$, a sharp line should emerge near $1.69$~eV upon cooling---a qualitative signature insensitive to defect density and collection efficiency, and thus the most direct test of our prediction. Once the ZPL can be identified, it can be selectively enhanced in an optical cavity~\cite{Froch2021}, where the Purcell effect acts on the coherent line alone and redistributes emission from the phonon sideband into it. This is the route by which stacking engineering turns a technologically mature but dim defect into a usable single-spin--photon interface. More broadly, our study underscores the potential of stacking engineering in two-dimensional materials as a general strategy for tailoring defect-based quantum systems.

\section{Methods}

We employed two complementary approaches---density functional theory (DFT) and many-body perturbation theory (MBPT) within the GW approximation---to describe the electronic structure of the V$_\text{B}^{-}$ defect at different levels of theory. For the DFT calculations, a plane-wave basis set with a cutoff of 450~eV and projector augmented-wave (PAW) atomic potentials~\cite{blochl1994projector} were used as implemented in VASP version 5.4.1~\cite{kresse1994ab, kresse1996efficient}. The convergence criteria for total energy and atomic forces were set to $10^{-4}$~eV and 0.01~eV/\AA, respectively. Structural relaxations, excited-state calculations within constrained occupation DFT (or $\Delta$SCF method)~\cite{gali2009theory}, hyperfine tensors~\cite{Szasz_2013}, and spin--spin zero-field splitting (ZFS) tensors~\cite{Bodrog_2013} as implemented by Martijn Marsman were obtained using the Heyd--Scuseria--Ernzerhof (HSE) hybrid functional~\cite{heyd2003hybrid} with a 0.32 fraction of exact exchange and a screening parameter of 0.2~\AA$^{-1}$. To eliminate spin contamination in the spin--spin ZFS tensor, we applied the correction scheme of Ref.~\citenum{biktagirov2020spin}. The supercell model consisted of a $9\times9\times1$ 486-atom rBN supercell embedding a single negatively charged boron vacancy. The optimized interlayer distance was 3.31~\AA, obtained using the DFT-D3 dispersion correction method of Grimme~\cite{grimme2010consistent}. Optically detected magnetic resonance (ODMR) spectra were simulated with the EasySpin package version 6.0.0~\cite{stoll2006easyspin} running under MATLAB R2023b, using the \texttt{pepper} routine in hybrid mode; the full input script is reproduced in Supplementary Note~4. The Jahn--Teller Hamiltonian was solved with our \texttt{jahn-teller-dynamics} Python package~\cite{Toth2025}.

Since excitation spectra based purely on DFT energy levels are not always reliable for comparison with experiment~\cite{attaccalite2011coupling}, we also employed MBPT within the GW approximation and the Bethe--Salpeter equation (BSE) to predict the optical properties of both defects and the host lattice. MBPT explicitly accounts for electron--hole interactions in the excited state, where the excited electron experiences the Coulomb potential of the remaining hole; these correlated excitations are described by a two-particle Green's function that fulfills the BSE~\cite{onida2002electronic}. Due to the high computational cost of MBPT, we used a smaller $6\times6\times1$ 216-atom supercell. We verified explicitly that this cell is converged for the quantities of interest: repeating the $\Delta$SCF excited-state calculation in $N\times N\times1$ ($N=5,6,7,9$) and $6\times6\times N$ ($N=1,2,3$) supercells changes the excitation energies by less than 20~meV between the $6\times6\times1$ and the production $9\times9\times1$ cell, and by less than 50~meV over the whole series, while the transition dipole moments vary by less than 20\% (Supplementary Note~6 and Supplementary Figure~4). We emphasize that the long-range $1/r$ finite-size errors that plague charged-defect formation energies largely cancel here, because the ground and excited states carry the same charge; the residual defect--defect coupling is of much shorter range (exponential in the orbital overlap, or $1/r^3$ dipole--dipole). For bulk pristine rBN, our GW+BSE calculations yielded a band gap of 5.82~eV at the $K$ point, in good agreement with experiment~\cite{qi2024stacking, sponza2018direct}.

Quasiparticle energies were obtained within the evGW$_0$ scheme, starting from HSE DFT results as input. This method provides an efficient and accurate approximation to the GW formalism. Quasiparticle wave functions and energies were updated until convergence was reached after four iterations. The plane-wave basis for the orbitals was kept at the 450~eV cutoff used in the DFT step, which exceeds the largest recommended cutoff of the B and N PAW potentials; the response function was expanded up to \texttt{ENCUTGW}~$=270$~eV, i.e. the VASP default of $2/3$ of the orbital cutoff, which is the standard converged choice for $sp^2$ BN. For the self-energy calculations, 6500 bands were included, corresponding to more than 30 times the number of occupied states of the 216-atom supercell. We note that the quantities we report are energy \emph{differences} between states described in one and the same basis, which converge substantially faster with respect to both the cutoff and the number of bands than the absolute quasiparticle energies do. Brillouin-zone sampling was restricted to the $\Gamma$ point, which proved convergent for both DFT and GW+BSE calculations. The defect-level energy differences obtained from GW were consistent with those from HSE (see Supplementary Note~5), validating their reliability for subsequent BSE optical property calculations.

\section*{Data availability}

The data supporting the findings of this study are computational. All quantities underlying the reported results---optimized supercell geometries, Kohn--Sham and quasiparticle eigenvalues, BSE excitation energies and transition dipole moments, partial Huang--Rhys factors and vibrational frequencies, and the hyperfine and zero-field-splitting tensors---are reported in full in the main text and in the Supplementary Information, which together constitute the minimum dataset needed to interpret, verify and extend this work. The underlying raw electronic-structure output files (VASP \texttt{OUTCAR}, \texttt{vasprun.xml}, \texttt{POSCAR} and \texttt{CONTCAR} files) are archived on the institutional storage of the HUN-REN Wigner Research Centre for Physics and are available from the corresponding author (A.G., gali.adam@wigner.hun-ren.hu), who will respond to requests within four weeks. No restrictions apply to their use. No experimental datasets were generated or analysed during the current study.

\section*{Code availability}

No custom code was used to generate the results of this study beyond the items listed here. The electronic-structure calculations were performed with the commercial plane-wave code VASP version 5.4.1, which is available under licence from the VASP Software GmbH; the zero-field-splitting module used within it is the standard implementation by Martijn Marsman distributed with that release, modified only by removing the $1/[S(2S-1)]$ prefactor for the broken-symmetry singlet configuration, as described in Supplementary Note~3. The optically detected magnetic resonance spectra were simulated with EasySpin version 6.0.0, an open-source package freely available at https://easyspin.org; the complete input script is reproduced verbatim in Supplementary Note~4, so that the simulations can be repeated without any additional code. The photoluminescence lineshape was obtained with our own Huang--Rhys implementation, described in Ref.~\citenum{thiering2017ab} and based on the theory of Alkauskas and co-workers~\cite{alkauskas2014first}. That code is not distributed publicly but is available from the corresponding author on request; it does not constitute a key advance of the present paper, since the advance is the physical result rather than the software, and it is fully documented in the cited publication. The $e\otimes E$ Jahn--Teller Hamiltonian was solved with the \texttt{jahn-teller-dynamics} Python package, which is freely available from the public repository linked in the code paper of Ref.~\citenum{Toth2025}.

\section*{Acknowledgements}

A.G.\ acknowledges access to high-performance computational resources provided by KIF\"U (Governmental Agency for IT Development, Hungary). G.T.\ acknowledges the support from the J\'anos Bolyai Research Scholarship of the Hungarian Academy of Sciences. N.E.\ acknowledges the scholarship support from the Iranian Ministry of Science, Research, and Technology.

\section*{Funding}

A.G.\ discloses support for the research of this work from the European Commission [SPINUS, Grant No.\ 101135699] and the European Commission [QuSPARC, Grant No.\ 101186889]. G.T.\ discloses support for the research of this work from the National Office of Research, Development and Innovation of Hungary (NKFIH) [STARTING Grant No.\ 150113]. N.E.\ and I.A.S.\ disclose support for the research of this work from the Iranian National Science Foundation (INSF) [Project No.\ 4021945].

\section*{Author contributions}

N.E.\ and G.T.\ performed the computations. The analysis of the results was carried out by N.E., supervised by I.A.S., G.T., and A.G. A.G.\ conceived and supervised the project. All authors contributed to manuscript writing.

\section*{Competing interests}

The authors declare no competing interests.

\bibliography{references}

\clearpage
\onecolumngrid

\setcounter{equation}{0}
\setcounter{figure}{0}
\setcounter{table}{0}
\renewcommand{\theequation}{S\arabic{equation}}
\renewcommand{\thefigure}{\arabic{figure}}
\renewcommand{\thetable}{\arabic{table}}
\renewcommand{\figurename}{Supplementary Figure}
\renewcommand{\tablename}{Supplementary Table}

\part*{\begin{center}Supplemental Material\end{center}}

\section*{Supplementary Note 1: Application of Huang-Rhys theory}

Within the Huang-Rhys (HR) theory, one takes the following approximations:
\begin{itemize}
\item Both the ground and excited states exhibit the same parabolic confinement
potential with the same set of vibration modes: ($\hbar\omega_{i}$),
where $i=1\dots(3N-3)$ spans all vibration modes of a rBN supercell
hosting $N$ atoms.
\item The electron-phonon coupling manifests itself only by that these two adiabatic potential energy surfaces (APES) are shifted from each other by a fixed amount by $2\alpha_{i}$. 
\end{itemize}
Optical transitions are accompanied by phonon sidebands. The corresponding lineshape can be described within the model of two displaced quantum harmonic oscillators. Because the minima of the two potential energy surfaces are displaced relative to each other, not only the zero-phonon-line (ZPL) transition is allowed, but transitions involving different phonon quanta also acquire finite intensity due to the nonzero overlap of their vibrational wavefunctions. The square of this dimensionless displacement is known as the partial Huang--Rhys (HR) factor, $S_i = 2\alpha_i^2$, which characterizes the strength of the electron--phonon coupling for each vibrational mode. Accordingly, the potential energy surfaces of the ground and excited electronic states, $V_{-}(x_i)$ and $V_{+}(x_i)$, can be written in terms of the dimensionless normal coordinates $x_i$ as follows:

\begin{equation}
V_{\mp}(x_{i})=\mp\frac{\Delta}{2}+\sum_{i}\Bigl(\frac{\hbar\omega_{i}}{2}x_{i}^{2}\mp f_{i}x_{i}\Bigr)=\mp\frac{\Delta}{2}+\sum_{i}\frac{\hbar\omega_{i}}{2}\Bigl(x_{i}\mp\frac{f_{i}}{\hbar\omega_{i}}\Bigr)^{2}-\frac{f_{i}^{2}}{(\hbar\omega_{i})^{2}}\text{,}
\end{equation}
where $\Delta$ is the bare electronic excitation energy. The quantity $f_i$ is the derivative of the total energy with respect to the normal coordinate, i.e., the generalized force experienced by the system upon electronic excitation. By completing the square, one finds that the dimensionless displacement is given by $\alpha_i = f_i / (\hbar \omega_i)$. In \emph{ab initio} calculations, however, the static displacement is typically obtained in \AA\ units, which allows the shift in generalized coordinates to be computed as follows~\cite{alkauskas2014first}:

Next, the generalized coordinates $q_i$ provide a convenient way to define the Huang--Rhys factors $S_i$ (or, equivalently, the dimensionless displacements $\alpha_i$):
\begin{equation}
S_i=\frac{\omega_i q_i^2}{2\hbar}
\quad\text{or}\quad
2\alpha_i=\sqrt{\frac{\omega_i}{\hbar}}\,q_i \, .
\end{equation}
These quantities allow the zero-temperature lineshape function to be written as follows (see the Supplementary Material of Ref.~\cite{PhysRevB.88.165202} for details):
\begin{equation}
F(\hbar\omega,T=0\,\mathrm{K})
=
\sum_{i,\nu_i}
\left|\langle 0_i^{(g)} | \nu_i^{(e)} \rangle\right|^2
\delta\!\left(\hbar\omega-\sum_{\nu_i}\nu_i\hbar\omega_i\right)
=
e^{-S_{\mathrm{tot}}}
\sum_{n=0}^{\infty}
\frac{S_{\mathrm{tot}}^n}{n!}\,
I_n(\hbar\omega)\, .
\end{equation}
Here, $|\nu_i^{(e)}\rangle$ denotes the vibrational wavefunction of the excited state with $\nu_i$ phonon quanta, whereas $\langle 0_i^{(g)}|$ denotes the vibrational wavefunction of the ground state with zero phonon quanta. The functions $I_n(\hbar\omega)$ are defined recursively by convolution,
\begin{equation}
I_n(\hbar\omega)
=
\int d(\hbar\omega^\prime)\,
I_{n-1}(\hbar\omega^\prime)\,
S(\hbar\omega-\hbar\omega^\prime)\, .
\end{equation}
Finally,
\begin{equation}
S(\hbar\omega)=I_1(\hbar\omega)=\sum_i S_i\,\delta(\hbar\omega-\hbar\omega_i)
\end{equation}
is the Huang--Rhys spectral function, which determines the first phonon sideband, whereas the zero-phonon line is given by
\begin{equation}
I_0(\hbar\omega)=\delta(\hbar\omega)\, ,
\end{equation}
and the total Huang--Rhys factor is
\begin{equation}
S_{\mathrm{tot}}=\sum_i S_i\, .
\end{equation}

At finite temperature ($T>0$), it is convenient to use the generating-function formalism based on the Fourier transform pair $S(\hbar\omega)\leftrightarrow S(t)$. The generating function is then defined as
\begin{equation}
G(t)=e^{S(t)-S(0)}\, ,
\end{equation}
where
\begin{equation}
S(t)
=
\int_{-\infty}^{+\infty}
\left[
n(\hbar\omega)e^{+i\omega t}
+
\bigl(n(\hbar\omega)+1\bigr)e^{-i\omega t}
\right]
S(\hbar\omega)\,
d(\hbar\omega)\, ,
\end{equation}
and
\begin{equation}
n(\hbar\omega)=\left[\exp\!\left(\frac{\hbar\omega}{k_B T}\right)-1\right]^{-1}
\end{equation}
is the Bose--Einstein occupation factor for the vibrational quanta. The temperature-dependent lineshape function is then given by
\begin{equation}
\boxed{
F(\hbar\omega,T)
=
\frac{1}{2\pi}
\int_{-\infty}^{+\infty}
G(t)\,e^{i\omega t-\gamma|t|}\,dt
}
\end{equation}
where $\gamma$ may be chosen to reproduce the experimental broadening of the ZPL.

One must still distinguish between the cases of photoluminescence (PL) and absorption (A):
\begin{equation}
I_{\mathrm{PL}}(\hbar\omega,T)
=
(\hbar\omega)^3\,
F(\Delta_{\mathrm{ZPL}}-\hbar\omega,T)
\quad\text{and}\quad
I_{\mathrm{A}}(\hbar\omega,T)
=
(\hbar\omega)\,
F(\Delta_{\mathrm{ZPL}}+\hbar\omega,T)\, ,
\end{equation}
where $\Delta_{\mathrm{ZPL}}$ is the zero-phonon transition energy. Thus, the PL and absorption lineshapes are weighted by factors of $(\hbar\omega)^3$ and $(\hbar\omega)^1$, respectively.

We additionally note that the Stokes shift, i.e., the difference between the vertical absorption and vertical emission energies, can be written as follows~\cite{de_Jong_2015}:
\begin{equation}
E_{\mathrm{Stokes}}
=
\underbrace{V_{+}(\alpha_i)-V_{-}(\alpha_i)}_{\mathrm{vertical\ absorption}}
-
\underbrace{\bigl[V_{+}(-\alpha_i)+V_{-}(-\alpha_i)\bigr]}_{\mathrm{vertical\ emission}}
=
\left(\Delta+\sum_i \hbar\omega_i S_i\right)
-
\left(\Delta-\sum_i \hbar\omega_i S_i\right)
\approx
2S_{\mathrm{tot}}\hbar\langle\omega_i\rangle \, ,
\end{equation}
where $\langle\omega_i\rangle$ denotes the $S_i$-weighted mean vibrational frequency,
\begin{equation}
\langle\omega_i\rangle=\frac{\sum_i S_i\omega_i}{S_{\mathrm{tot}}}\, .
\end{equation}

\begin{figure}[h]
  \includegraphics[width=0.45\textwidth]{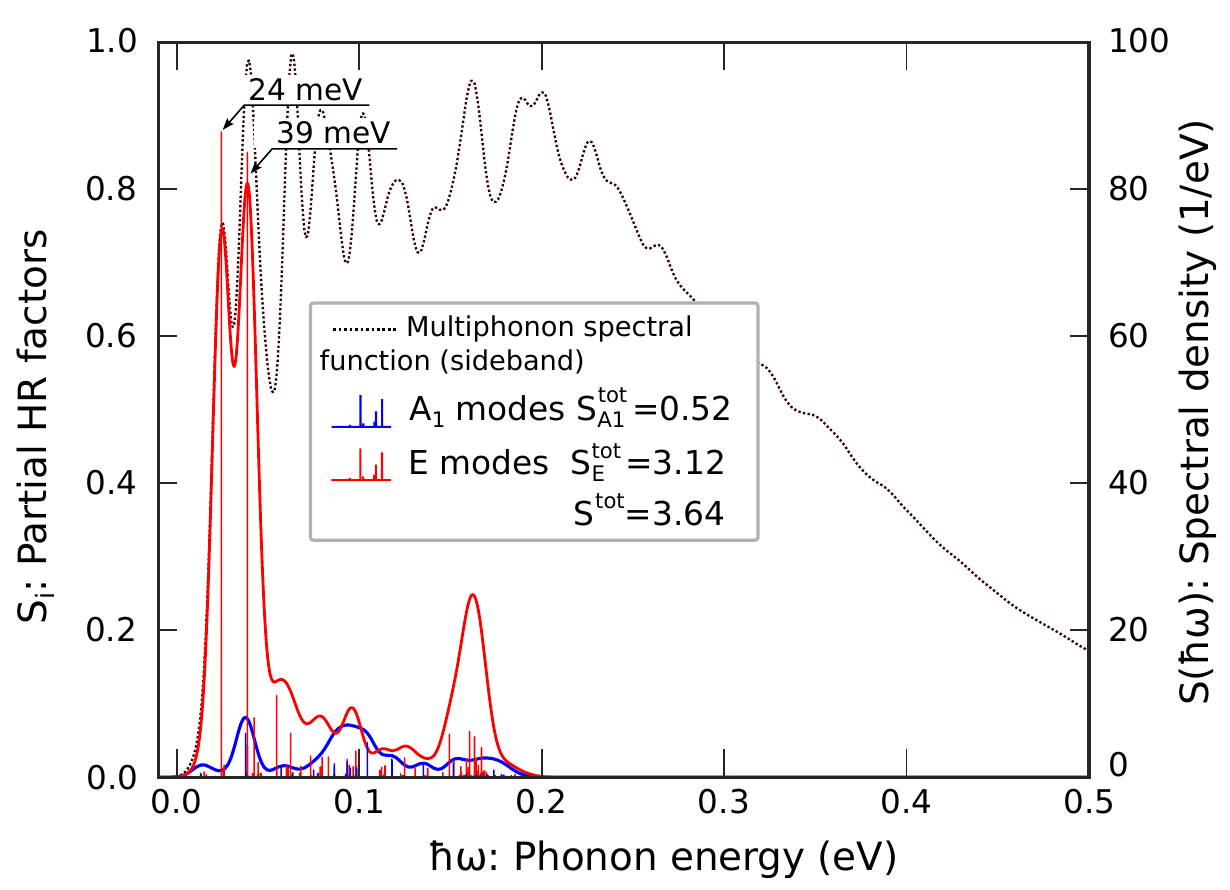}
    \caption{\label{fig:SU} Partial Huang-Rhys factors individually distributed into totally symmetric $A_1$ and Jahn-Teller active $E$ modes at 0 K temperature.}
\end{figure}

\section*{Supplementary Note 2: Hyperfine tensors}

Calculated values of the hyperfine coupling constants for nitrogen and boron nuclear spins are presented in Supplementary Tables~1 and 2, respectively, for the most significant neighboring sites of VB, see Supplementary Figure~2(a). Since the hyperfine interaction with farther nuclear spins is much weaker, we only consider the nearest nuclear spins. The spin density of the excited state shows a reduction of symmetry as shown in Supplementary Figure~2(b).
\begin{figure}[h]
  \includegraphics[width=0.45\textwidth]{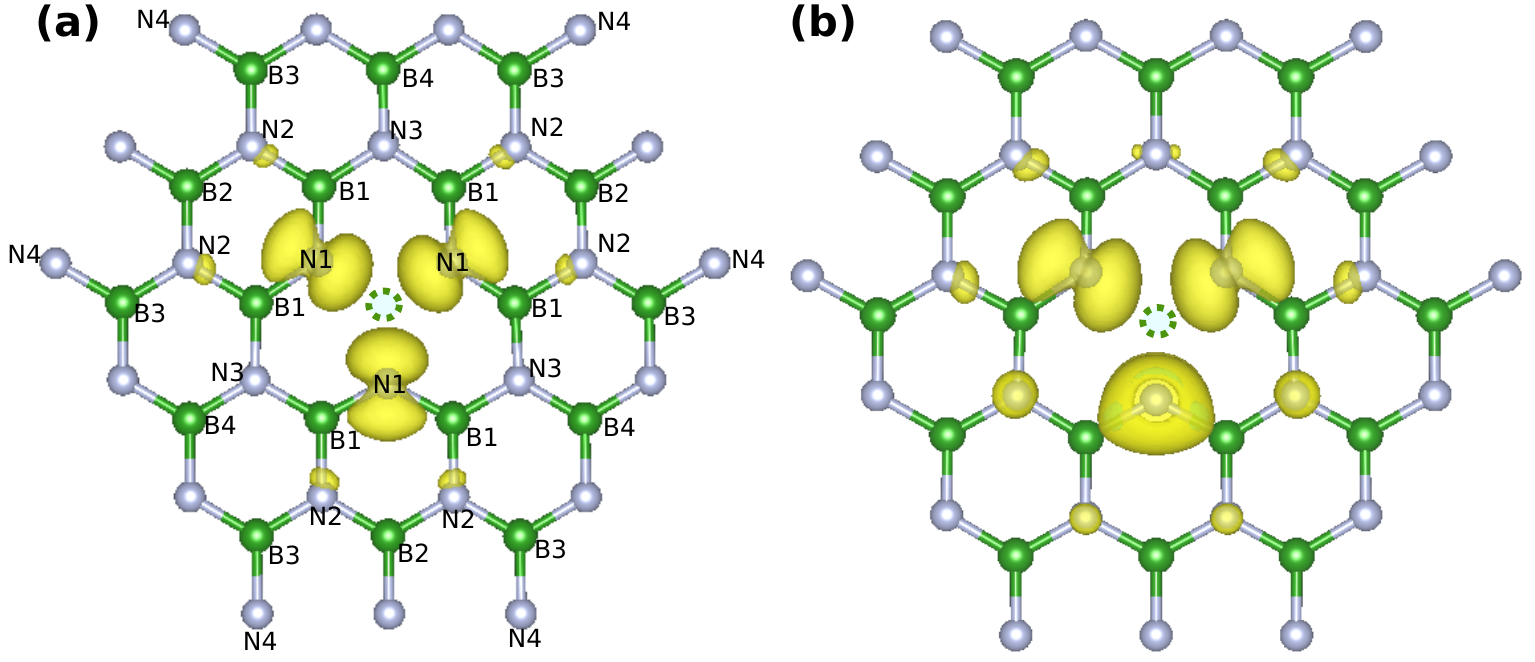}
    \caption{\label{fig: spin-density} Calculated spin density. (a) Ground state. (b) Excited state. The vacant boron site located in the middle layer of rBN is depicted as a semitransparent green ball.}
\end{figure}

\begin{table*}[h]
\caption{\label{tab:hyperfine_parameters_N} \textit{Ab initio} hyperfine parameters for $^{14}$N nuclei $I=1$ nuclear spins in the negatively charged boron-vacancy in rBN. All hyperfine values are in MHz unit.}
 \begin{ruledtabular}
 \begin{tabular}{c|cccc}
         site & $A_{xx}$ &  $A_{yy}$ & $A_{zz}$ & $A_{xy}$\\
        \hline
        $N_1$ & 80.566 & 57.865 & 48.304 & 19.670  \\
        $N_1$ & 80.584 & 57.873 & 48.329 & -19.669 \\
        $N_1$ & 46.505 & 91.924 & 48.303 & 0.005  \\
        $N_2$ & 6.699 & 4.484 & 4.372 & 0.594  \\
        $N_2$ & 5.553 & 5.628 & 4.371 & 1.256  \\
        $N_2$ & 6.694 & 4.480 & 4.364 & -1.255  \\
        $N_2$ & 6.694 & 4.480 & 4.364 & -0.593  \\
        $N_2$ & 4.522 & 6.659 & 4.371 & 0.660 \\
        $N_2$ & 4.524 & 6.659 & 4.372 & -0.662  \\
        $N_3$ & 0.830 & 0.172 & -0.398 & 0.000  \\
        $N_3$ & 0.335 & 0.665 & -0.399 & 0.286  \\
        $N_3$ & 0.336 & 0.666 & -0.398 & -0.285  \\
        $N_4$ & 0.315 & 0.367 & 0.246 & 0.101  \\
        $N_4$ & 0.439 & 0.237 & 0.242 & 0.029  \\
        $N_4$ & 0.262 & 0.414 & 0.242 & -0.073  \\
        $N_4$ & 0.266 & 0.415 & 0.246 & 0.073  \\
        $N_4$ & 0.442 & 0.240 & 0.247 & -0.029  \\
        $N_4$ & 0.316 & 0.367 & 0.247 & -0.102  \\
    \end{tabular}
 \end{ruledtabular}  
\end{table*}

\begin{table*}[h]
\caption{\label{tab:hyperfine_parameters_B}  \textit{Ab initio} hyperfine parameters for $^{11}$B nuclei with $I=3/2$ nuclear spins in the negatively charged boron-vacancy in rBN. All hyperfine values are in MHz unit.}
 \begin{ruledtabular}
 \begin{tabular}{c|cccc}
         site & $A_{xx}$ &  $A_{yy}$ & $A_{zz}$ & $A_{xy}$\\
        \hline
        $B_1$ & -1.740 & 5.358 & -4.326 & 0.899 \\
        $B_1$ & -1.740 & 5.358 & -4.326 & -0.899  \\
        $B_1$ & 2.805 & 0.813 & -4.326 &  3.523 \\
        $B_1$ & 2.805 & 0.813 & -4.326 &  -3.523 \\
        $B_1$ & 4.362 & -0.744 & -4.326 & 2.624 \\
        $B_1$ & 4.362 & -0.744 & -4.326 & -2.624  \\
        $B_2$ & -1.213 & 0.167 & -1.651 & 0.000  \\
        $B_2$ & -0.178 & -0.868 & -1.651 & 0.597  \\
        $B_2$ & -0.178 & -0.868 & -1.651 & -0.597  \\
        $B_3$ & 2.920 & 2.057 & 1.725 & -0.446  \\
        $B_3$ & 1.885 & 3.090 & 1.724 &  0.151 \\
        $B_3$ & 1.886 & 3.090 & 1.725 & -0.151  \\
        $B_3$ & 2.920 & 2.057 & 1.725 &  0.446 \\
        $B_3$ & 2.658 & 2.317 & 1.724 & 0.597  \\
        $B_3$ & 2.659 & 2.318 & 1.725 & -0.597  \\
        $B_4$ & -0.977 & -0.272 & -1.196 & 0.000  \\
        $B_4$ & -0.447 & -0.800 & -1.195 & -0.306  \\
        $B_4$ & -0.448 & -0.801 & -1.196 & 0.305  \\
    \end{tabular}
 \end{ruledtabular}  
\end{table*}

\section*{Supplementary Note 3: Determination of ZFS tensors}
\subsubsection*{Determination of ZFS tensors}

The zero-field-splitting (ZFS) tensor for a two-particle wavefunction
\begin{equation}
|\psi_i \psi_j\rangle
=
\frac{\psi_i(r_1)\psi_j(r_2)-\psi_j(r_1)\psi_i(r_2)}{\sqrt{2}}
\end{equation}
constructed from orbitals $\psi_i$ and $\psi_j$ with the same spin projection (e.g.\ both $\uparrow$) can be calculated as follows, including exchange effects~\cite{PhysRevB.77.035119,Bodrog_2013}:
\begin{equation}
\begin{split}D^{(ij)}=\frac{3}{2}D_{zz}^{(ij)}=\left\langle \psi_{i}\psi_{j}\left|\frac{3}{2}f_{zz}\right|\psi_{i}\psi_{j}\right\rangle = & \frac{3}{2}\biggl(\langle\psi_{i}(r_{1})\psi_{j}(r_{2})|f_{zz}(r_{1}-r_{2})|\psi_{i}(r_{1})\psi_{j}(r_{2})\rangle\\
 & \;-\langle\psi_{i}(r_{1})\psi_{j}(r_{2})|f_{zz}(r_{1}-r_{2})|\psi_{j}(r_{1})\psi_{i}(r_{2})\rangle\biggr)\,,
\end{split}
\end{equation}
where
\begin{equation}
f_{ab}(r)=\frac{r^2\delta_{ab}-3r_a r_b}{r^5}
\end{equation}
is the magnetic dipole--dipole kernel, with $a,b\in\{x,y,z\}$ denoting Cartesian coordinates and $r=\sqrt{x^2+y^2+z^2}$.

At this point, we consider a $C_{3v}$-symmetric ZFS tensor, which has only one independent parameter:
\begin{equation}
D_{ab}
=
\begin{pmatrix}
D_{xx} & D_{xy} & D_{xz}\\
D_{xy} & D_{yy} & D_{yz}\\
D_{xz} & D_{yz} & D_{zz}
\end{pmatrix}
=
\frac{1}{3}
\begin{pmatrix}
-D & 0 & 0\\
0 & -D & 0\\
0 & 0 & 2D
\end{pmatrix}.
\end{equation}
Therefore, it is sufficient to evaluate the $D_{zz}$ tensor element. For simplicity, the physical constants $\mu_B$ and $g$ are omitted.

This simplification becomes especially relevant for the $|^3E\rangle$ excited state, where orbital degeneracy can in principle induce off-diagonal elements in $D_{ab}$. However, these effects are expected to be absent even at low temperatures for V$_\text{B}^-$ in rBN because of strong electron-phonon interaction (see the small Ham reduction factor). As a consequence, only the splitting between the $m_S=\pm1$ and $m_S=0$ sublevels remains, which can then be parametrized by a single ZFS constant $D$.

More generally, for an $N$-electron system, the ZFS tensor can be evaluated as a sum over occupied spin orbitals. For example, for the $|^3A_2^{\mathrm{gnd}}\rangle$ state of $\mathrm{V_B^-}$,
\begin{equation}
D_{\mathrm{gnd}}
=
\frac{1}{S(2S-1)}
\sum_{i,j}^{\mathrm{occ.}} D^{(ij)}
\, ,
\end{equation}
where the sum runs over all occupied orbitals in both spin channels. In the present case,
\begin{equation}
i,j \in
\{
\mathrm{VB\ orbitals},
a_1^{\uparrow},a_1^{\downarrow},
a_1^{\star\uparrow},a_1^{\star\downarrow},
e_x^{\uparrow},e_x^{\downarrow},
e_y^{\uparrow},e_y^{\downarrow},
e_x^{\star\downarrow},e_y^{\star\downarrow}
\},
\end{equation}
as depicted in Fig.~1b of the main text. We emphasize that the sum must also include all valence-band (VB) orbitals. By contrast, the orbitals $\{e_x^{\star\uparrow},e_y^{\star\uparrow}\}$ are unoccupied and therefore do not enter the sum. For triplet states ($S=1$), the prefactor $1/[S(2S-1)]$ is equal to unity.

We compute the ZFS tensor using the implementation in VASP 5.4.1 as implemented by Martijn Marsman. For the two-hole configuration
\begin{equation}
|^3A_2^{\mathrm{gnd},m_S=+1}\rangle
=
|e_x^{\star\uparrow} e_y^{\star\uparrow}\rangle,
\end{equation}
accessible within our DFT approach, we obtain
\begin{equation}
D_{\mathrm{ZFS}}\!\left[|e_x^{\star\uparrow} e_y^{\star\uparrow}\rangle\right]
=
2627~\mathrm{MHz}.
\end{equation}

As noted in the main text, we employ the spin-decontamination scheme proposed in Ref.~\citenum{biktagirov2020spin}, which requires evaluation of the ZFS for the broken-symmetry singlet configuration:
\begin{equation}
D\!\left[|e_y^{\star\uparrow}e_x^{\star\downarrow}\rangle\right]
=
-4250~\mathrm{MHz}.
\end{equation}
In this case, the prefactor $1/[S(2S-1)]$ must be removed from the VASP source code to avoid division by zero.

The corrected ground-state ZFS is then obtained by averaging the ferromagnetic and antiferromagnetic solutions, taking into account the sign change of the latter:
\begin{equation}
D_{\mathrm{theory}}^{\mathrm{gnd}}
=
\frac{1}{2}
\left(
D\!\left[|e_x^{\star\uparrow}e_y^{\star\uparrow}\rangle\right]
-
D\!\left[|e_x^{\star\uparrow}e_y^{\star\downarrow}\rangle\right]
\right)
=
\frac{1}{2}\bigl(2627~\mathrm{MHz}+4250~\mathrm{MHz}\bigr)
=
3438~\mathrm{MHz}
\, .
\end{equation}
This value is remarkably close to the experimentally observed room-temperature value,
\begin{equation}
D_{\mathrm{expt.}}^{\mathrm{gnd}}\approx 3.45~\mathrm{GHz},
\end{equation}
which supports the plausibility of the correction scheme for $\mathrm{V_B^-}$.

\subsubsection*{Tentative derivation of the ZFS for the $|^3E\rangle$ excited multiplet}

We first construct the symmetry-adapted wavefunctions for the ground and excited multiplets:
\begin{equation}
\begin{array}{ccc}
|^3A_2^{\mathrm{gnd}}\rangle=|e_x^\star e_y^\star\rangle \, , & & \\[4pt]
|^3A_1\rangle=\bigl(|e_x e_x^\star\rangle+|e_y e_y^\star\rangle\bigr)/\sqrt{2}\, ,
& \qquad &
|^3E_x\rangle
=
a|a_1 e_x^\star\rangle
+
b\bigl(-|e_x e_x^\star\rangle+|e_y e_y^\star\rangle\bigr)
+
c|a_1^\star e_x^\star\rangle \, , \\[6pt]
|^3A_2\rangle=\bigl(|e_x e_y^\star\rangle-|e_y e_x^\star\rangle\bigr)/\sqrt{2}\, ,
& \qquad &
|^3E_y\rangle
=
a|a_1 e_y^\star\rangle
+
b\bigl(|e_x e_y^\star\rangle+|e_y e_x^\star\rangle\bigr)
+
c|a_1^\star e_y^\star\rangle \, .
\end{array}
\end{equation}
For simplicity, the $\uparrow$ spin labels are omitted. According to our GW+BSE results, the expansion coefficients are $a^2=0.57$, $2b^2=0.32$, and $c^2=0.07$, as shown in Fig.~1c of the main text. Here, $|^3E_{x,y}\rangle$ refers to the lower $|^3E\rangle$ multiplet. The $|^1A_1\rangle$ level lies well above the upper $|^3E\rangle$ state, is optically inactive, and appears at 2.35~eV.

Because we cannot evaluate the ZFS directly on top of the GW+BSE wavefunctions, we instead use $\Delta$SCF calculations. In this approach, we obtain the following two values entering the decontamination correction:
\begin{equation}
D_{\Delta \mathrm{SCF}}^{\mathrm{ex.}}
=
\frac{1}{2}
\left(
D\!\left[\Psi_{\Delta \mathrm{SCF}}^{\uparrow\uparrow}\right]
-
D\!\left[\Psi_{\Delta \mathrm{SCF}}^{\uparrow\downarrow}\right]
\right)
=
\frac{1}{2}
\bigl(-564~\mathrm{MHz}+3579~\mathrm{MHz}\bigr)
=
1508~\mathrm{MHz},
\end{equation}
where
\begin{equation}
\label{eq:HI}
\Psi_{\Delta \mathrm{SCF}}^{\uparrow\uparrow}
\approx
|aa_1^{\uparrow}+be_x^{\uparrow}+ca_1^{\star\uparrow};e_y^{\star\uparrow}\rangle
\end{equation}
and
\begin{equation}
\label{eq:OS}
\Psi_{\Delta \mathrm{SCF}}^{\uparrow\downarrow}
\approx
|aa_1^{\uparrow}+be_x^{\uparrow}+ca_1^{\star\uparrow};e_y^{\star\downarrow}\rangle
\end{equation}
are the single-Slater-determinant triplet and broken-symmetry singlet $\Delta$SCF solutions, respectively, approximately representing the $|^3E_y\rangle$ triplet. 

\begin{figure}[h]
  (a)\includegraphics[width=0.2\textwidth]{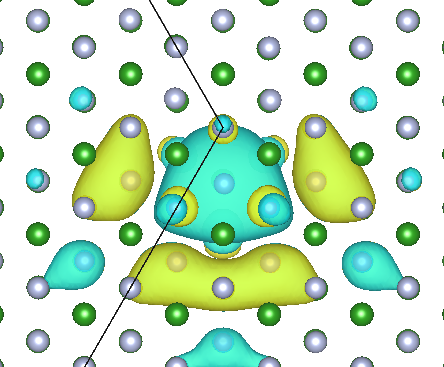}
  (b)\includegraphics[width=0.2\textwidth]{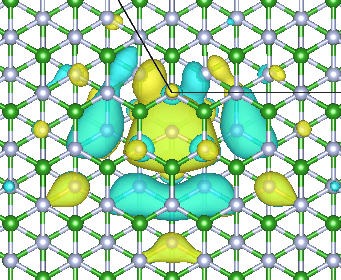}
    \caption{\label{fig: broken} Visualization of the broken-symmetry Kohn-Sham orbital $|aa_{1}^{\uparrow}+be_{x}^{\uparrow}+ca_{1}^{\star\uparrow}\rangle$ that of (a) $\Psi_{\Delta \mathrm{SCF}}^{\uparrow\uparrow}$ in Eq.~\eqref{eq:HI}  and (b) $\Psi_{\Delta \mathrm{SCF}}^{\uparrow\downarrow}$ in Eq.~\eqref{eq:OS}.}
\end{figure}

Thus, we identify the $\Delta$SCF component of $|^3E_y\rangle$ as
\begin{equation}
|^3E_y\rangle
=
\underbrace{|aa_1+ca_1^\star+be_x; e_y^\star\rangle}_{\sqrt{1-b^2}\,\Psi_{\Delta \mathrm{DFT}}^{\uparrow\uparrow}}
+
b\,|e_y e_x^\star\rangle
\, .
\end{equation}
For simplicity, we neglect the multiconfigurational contribution $b\,|e_y e_x^\star\rangle$ when estimating the excited-state ZFS.

\section*{Supplementary Note 4: Easyspin code}
We generated Fig.~3 of the main text by the following Easyspin code:

\begin{verbatim}
clear

Sys.S = 1; % two electrons spin with S=1/2
Sys.g = 2;  % isotropic g for two electron spins
%Sys.lw= 10.0;  % Gaussian Isotropic broadening
Sys.HStrain = [10 10 10];  %Gaussian FWHM Anisotropic broadening in all directions [MHz]
%(For hyperfine splittings)
Sys.D= [3450 10]; %MHz [D E]
Sys.A = [4.362 -0.744 -4.326; ... %11B hyperfine tensors: Axx, Ayy, Azz
         4.362 -0.744 -4.326; ... %11B.
         2.805  0.813 -4.326; ... %11B
         2.805  0.813 -4.326; ... %11B
        -1.74   5.358 -4.326; ... %11B
        -1.74   5.358 -4.326; ... %11B
        80.566 57.865 48.312; ... %14N
        46.506 91.925 48.304; ... %14N
        80.566 57.865 48.304];    %14N %MHz

Sys.Nucs = '11B, 11B, 11B, 11B, 11B, 11B, 14N, 14N, 14N';
Exp.Field = 9; %mT
Exp.Temperature = 4; %kelvin
Exp.mwRange = [3 4]; %GHz
%Exp.mwFreq = [];
Exp.CrystalSymmetry= 'C3v';
Exp.MolFrame = [0 0 0];
Exp.SampleFrame = [0 0 0]; %sample aligned with lab frame
%Exp.CrystalOrientation = [0 0 1];
Exp.nPoints = 1500;
%Exp.Range = [0 100];
%Opt.Method = 'perturb1';
%Opt.Verbosity = 1;

Opt.Method = 'hybrid';
Opt.HybridCoreNuclei = [7 8 9]; % these atoms treated exactly and others perturbatively

pepper(Sys,Exp,Opt);
\end{verbatim}

\section*{Supplementary Note 5: Comparison of evGW0 and DFT-HSE levels}

In Supplementary Table~\ref{tab:evGW} we list the quasiparticle and Kohn-Sham levels in the $\Gamma$-point for the 216-atom $6\times6\times1$ supercell. The results imply that the DFT-HSE spectrum is a good starting point for the excited state calculation by the $\Delta$SCF method.   

\begin{table}[h]
\caption{\label{tab:evGW}  The evGW0 quasiparticle (DFT-HSE) levels in the $\Gamma$-point are given in eV unit as obtained in 216-atom supercell.}
 \begin{ruledtabular}
 \begin{tabular}{cccccc}
 $\uparrow$ orb. & value (eV) & occ. & $\downarrow$ orb. & value (eV) & occ. \\ \hline
VBM & 0.89(1.13) & 1 & VBM & 0.89(1.13) & 1 \\
$a_{1}^{\uparrow}$ & 0.89(1.13) & 1 & $a_{1}^{\downarrow}$ & 1.53(1.71) & 1 \\
$e_{x}^{\uparrow}$ & 1.48(1.29) & 1 & $e_{x}^{\downarrow}$ & 1.47(1.72)  & 1 \\
$e_{y}^{\uparrow}$ & 1.48(1.29) & 1 & $e_{y}^{\downarrow}$ & 1.47(1.72)  & 1 \\
$a_{1}^{\star\uparrow}$ & 1.16(1.42) & 1 & $a_{1}^{\star\downarrow}$ & 2.32(1.81) & 1 \\
$e_{x}^{\star\uparrow}$ & 1.26(1.50) & 1 & $e_{x}^{\star\downarrow}$ & 5.56(5.70)  & 0 \\
$e_{y}^{\star\uparrow}$ & 1.26(1.50) & 1 & $e_{y}^{\star\downarrow}$ & 5.56(5.70) & 0 \\
CBM & 7.61(7.13) & 0 & CBM & 7.61(7.13) & 0 \\
\end{tabular}
 \end{ruledtabular}
\end{table}

\section*{Supplementary Note 6: Finite size convergence}

Both Reviewers asked whether the $6\times6\times1$ (216-atom) supercell employed in the GW+BSE calculations is sufficiently converged, given that the ground-state properties reported in the main text were obtained in a $9\times9\times1$ (486-atom) supercell. We address this here.

We first note a point of principle. The $1/r$ Coulomb finite-size errors that make charged defects notoriously slowly convergent enter the \emph{formation energy} and hence the charge transition levels, because there the charge state of the supercell changes. All quantities reported here---excitation energies, transition dipole moments, the phonon sideband, the ZFS and the hyperfine tensors---are instead evaluated within one and the same charge state, so the monopole and dipole image terms cancel between the initial and final state to leading order. What remains is the direct defect--defect coupling, which decays exponentially with the separation of the localized dangling-bond orbitals, and the residual dipole--dipole term, which decays as $1/r^3$. Convergence is therefore expected to be, and is found to be, much faster than for charge transition levels.

\begin{figure}[h]
  \includegraphics[width=0.9\textwidth]{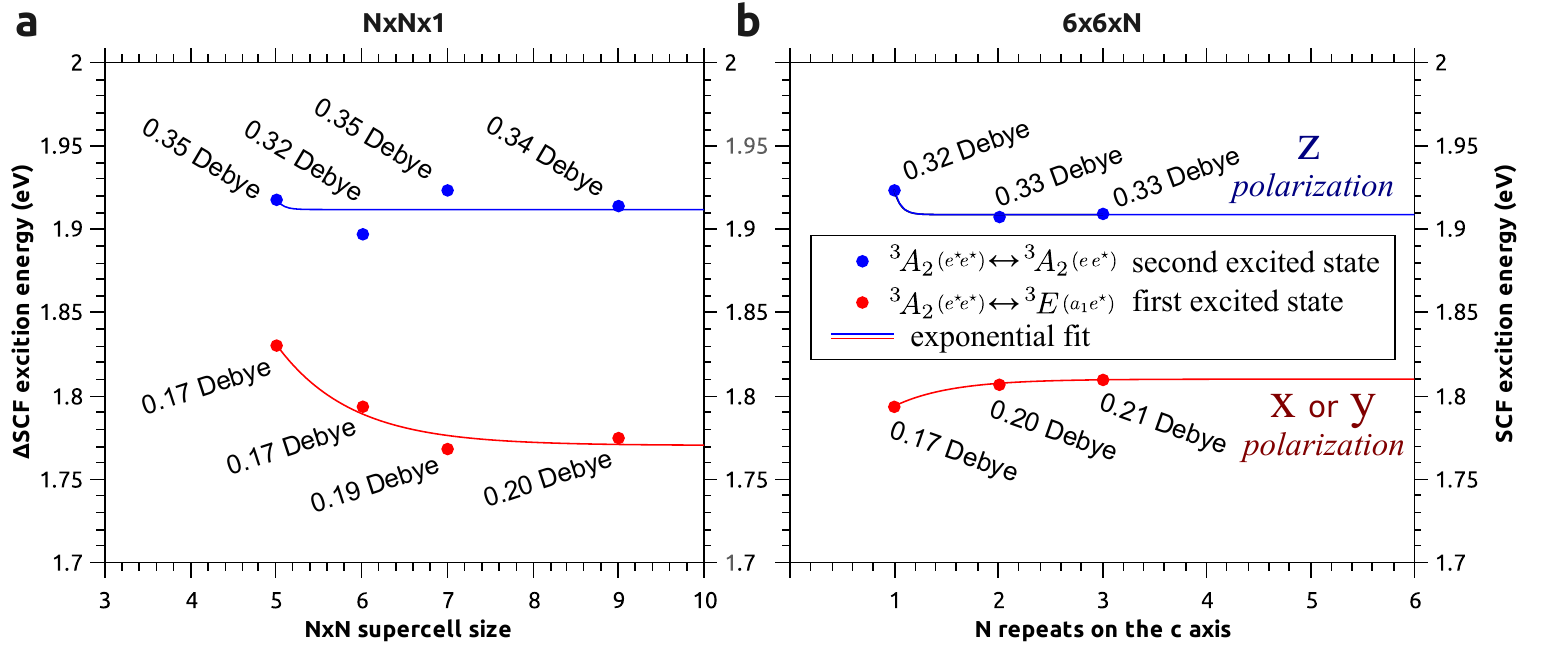}
 
    \caption{\label{fig:scaling} Finite-size convergence of the $\Delta$SCF excited states of V$_\text{B}^{-}$ in rBN. Red symbols, first excited state $^3A_2{}_{(e^\star e^\star)}\leftrightarrow{}^3E_{(a_1e^\star)}$, polarized perpendicular to the $C_3$ axis; blue symbols, second excited state $^3A_2{}_{(e^\star e^\star)}\leftrightarrow{}^3A_2{}_{(ee^\star)}$, polarized parallel to it. Solid lines are exponential fits to the asymptote. The number next to each point is the corresponding Kohn--Sham transition dipole moment, used here as a convergence metric only (see text).
    \textbf{a} In-plane enlargement of the BN layer, $N\times N\times1$ with $N=5$, 6, 7, 9, hosting 149, 215, 293 and 485 atoms, respectively.
    \textbf{b} Stacking of supercells along the $c$ axis, $6\times6\times N$ with $N=1$, 2, 3, hosting 215, 431 and 647 atoms, respectively.}
\end{figure}

To make this quantitative we repeated the constrained-occupation ($\Delta$SCF) HSE calculation of the two lowest excited states in six supercells, spanning both in-plane and out-of-plane directions (Supplementary Fig.~\ref{fig:scaling}). Because the same $\Delta$SCF protocol can be run in every cell, including the $6\times6\times1$ cell used for GW+BSE and the $9\times9\times1$ cell used for the production runs, this provides a direct, like-for-like test of the effect of the cell size on the excited state, which is exactly the quantity the Reviewers ask about.

\emph{In-plane extension} [Supplementary Fig.~\ref{fig:scaling}a; $N\times N\times1$ with $N=5$, 6, 7, 9 hosting 149, 215, 293 and 485 atoms]. Going from $6\times6\times1$ to $9\times9\times1$ changes the excitation energy of the first ($^3E$) excited state by $-18$~meV and that of the second excited state by $+16$~meV. Over the entire series the total spread is below 50~meV for the second excited state and about 60~meV for the first, with a clean exponential approach to the asymptote (solid lines are exponential fits).

\emph{Out-of-plane extension} [Supplementary Fig.~\ref{fig:scaling}b; $6\times6\times N$ with $N=1$, 2, 3 hosting 215, 431 and 647 atoms]. Doubling the cell along $c$ shifts the $\Delta$SCF excitation energies by only $\sim$20~meV, and tripling it by a further $\sim$2~meV. The rapid convergence reflects the fact that a single $6\times6\times1$ rBN supercell already contains three BN sheets, so the local ABC stacking environment of the vacancy---which is what breaks the mirror symmetry and drives the entire effect reported in this work---is fully represented.

The transition dipole moments, which enter the radiative rate quadratically, are given next to each data point in Supplementary Fig.~\ref{fig:scaling}. They vary between 0.17 and 0.21~D for the first excited state and between 0.32 and 0.35~D for the second, i.e. by less than 20\% across all cells, corresponding to less than a factor of 1.5 in the radiative rate. We emphasize that these are single-particle (Kohn--Sham) transition dipoles evaluated at the $\Delta$SCF level; they are used here purely as a \emph{convergence metric}. The absolute dipole moments quoted in the main text (Fig.~1d) are the excitonic dipoles from GW+BSE, which are larger because the correlated exciton is a coherent superposition of several electron--hole pairs ($a_1\to e^\star$, $e\to e^\star$ and $a_1^\star\to e^\star$, see Fig.~1c) whose contributions add constructively.

We conclude that the $6\times6\times1$ supercell reproduces the $9\times9\times1$ result to better than 20~meV in the excitation energy and better than 20\% in the transition dipole moment. These residual finite-size errors are far smaller than the $\sim$0.1~eV intrinsic accuracy of the underlying electronic-structure methods, and they are entirely negligible compared with the one- to two-orders-of-magnitude difference in radiative rate between rBN and hBN that constitutes the central claim of this work. The reason the larger $9\times9\times1$ cell was nevertheless used for the ground-state properties is not convergence of the excitation energy but the phonon sideband, the hyperfine and the ZFS tensors: the Huang--Rhys analysis requires a dense set of supercell vibrational modes and the hyperfine tensors require several shells of neighbors, both of which benefit from the larger cell, and both of which are affordable at the HSE level but not at the GW+BSE level.

\end{document}